\documentclass[conference]{IEEEtran}
\usepackage[hidelinks]{hyperref}

\usepackage[pdftex]{graphicx}

\ifCLASSINFOpdf
\else
\fi
\usepackage{amsmath}
\begin{document}
%
\title{Flow-based analysis of storage usage in a low-carbon European electricity scenario}



%
\author{\IEEEauthorblockN{
Bo Tranberg\IEEEauthorrefmark{1}\IEEEauthorrefmark{2},
Mirko Sch{\"a}fer\IEEEauthorrefmark{1},
Tom Brown\IEEEauthorrefmark{3},
Jonas H{\"o}rsch\IEEEauthorrefmark{3}\IEEEauthorrefmark{4} and
Martin Greiner\IEEEauthorrefmark{1}}
\IEEEauthorblockA{\IEEEauthorrefmark{1}Department of Engineering, Aarhus University, 8000 Aarhus C, Denmark. Email: bo@eng.au.dk}
\IEEEauthorblockA{\IEEEauthorrefmark{2}Danske Commodities A/S, 8000 Aarhus C, Denmark}
\IEEEauthorblockA{\IEEEauthorrefmark{3}Institute for Automation and Applied Informatics, Karlsruhe Institute of Technology, Germany}
\IEEEauthorblockA{\IEEEauthorrefmark{4}Frankfurt Institute for Advanced Studies, 60438~Frankfurt~am~Main, Germany}}


\maketitle

\begin{abstract}
The application of the flow tracing method to power flows in and out of storage units allows to analyse the usage of this technology option in large-scale interconnected electricity systems. We apply this method to a data-driven model of the European electricity network, which uses a techno-economic optimisation to determine generation and storage capacities and dispatch, assuming a 95\% reduction of CO$_2$ emission compared to 1990 levels. A flow-based analysis of the power inflow into the different storage technologies confirms the intuition that longer-term hydrogen storage is mainly utilised for wind, whereas short-term battery storage mostly receives inflow from solar power generation. The usage of storage technologies in general shows a local-but-global behaviour: Whereas on average the power outflow from these capacities is predominantly consumed locally inside the same node, when exported it is also transmitted over long distances as a global flexibility option for the entire system.
\end{abstract}


%
\IEEEpeerreviewmaketitle

\section{Introduction}
The European Union has set a target to reduce CO$_2$ emissions by 80-95\% in 2050 compared to 1990 levels \cite{roadmap2050}. Most scenarios for reaching this target rely on the large-scale integration of intermittent wind and solar power generation, which requires future investments in transmission and storage capacity to smooth the variable generation over large spatial distances and appropriate time scales. The seeming dichotomy of these two flexibility sources emphasises the need to understand their actual interplay in cost-efficient scenarios of a future low-carbon electricity system.



Energy system models often employ a global optimisation approach to derive cost optimal scenarios~\cite{pfenninger_energy_2014}. Even when all input data and modelling details are available, the complexities and interdependencies inherent to such models tend to impede a deeper understanding of the mechanisms at play in an optimal combination of resources and technologies. This in particular applies to the role of the electricity grid, given that the pooling nature of power transmission in general disguises the influence of individual nodes on the global flow pattern. In this context, the method of flow tracing has been shown to yield important insights. By following the path of partial power flows through the transmission network, this technique allows to connect the location of consumption with the location of generation, and to allocate power flows on transmission lines to exporters and importers~\cite{Bialek,Kirschen1997,Hoersch2018}. It has been proposed  for instance as a flow allocation method as part of the inter transmission system operator compensation (ITC) mechanism~\cite{itc}, or as the basis of a demand-side-oriented carbon emission allocation method~\cite{carbon-ft}. In the context of the system analysis of highly renewable electricity scenarios, flow tracing has been applied to allocate transmission capacities~\cite{Tranberg2015,Schaefer2017}, or as a technique to introduce flow-based nodal systems costs~\cite{Tranberg2018,eem17}. In this contribution we introduce an extended application of the generalised flow tracing method, which traces flows in and out of storage and is able to keep track of the originating source of generation. We apply this method to a low-carbon future European electricity scenario first presented in~\cite{Hoersch2017}.


This article is organised as follows: Section~\ref{sec:power} introduces the modelling of dispatch and investments in generation capacities of a low-carbon European electricity scenario. Section~\ref{sec:ft} reviews the flow tracing methodology and introduces the formulation for including storage facilities. Results are presented and discussed in Section~\ref{sec:res}, and Section~\ref{sec:con} presents the conclusions.

\section{Power system modelling}\label{sec:power}
\begin{figure*}[t]
\centering
\includegraphics[width=\textwidth]{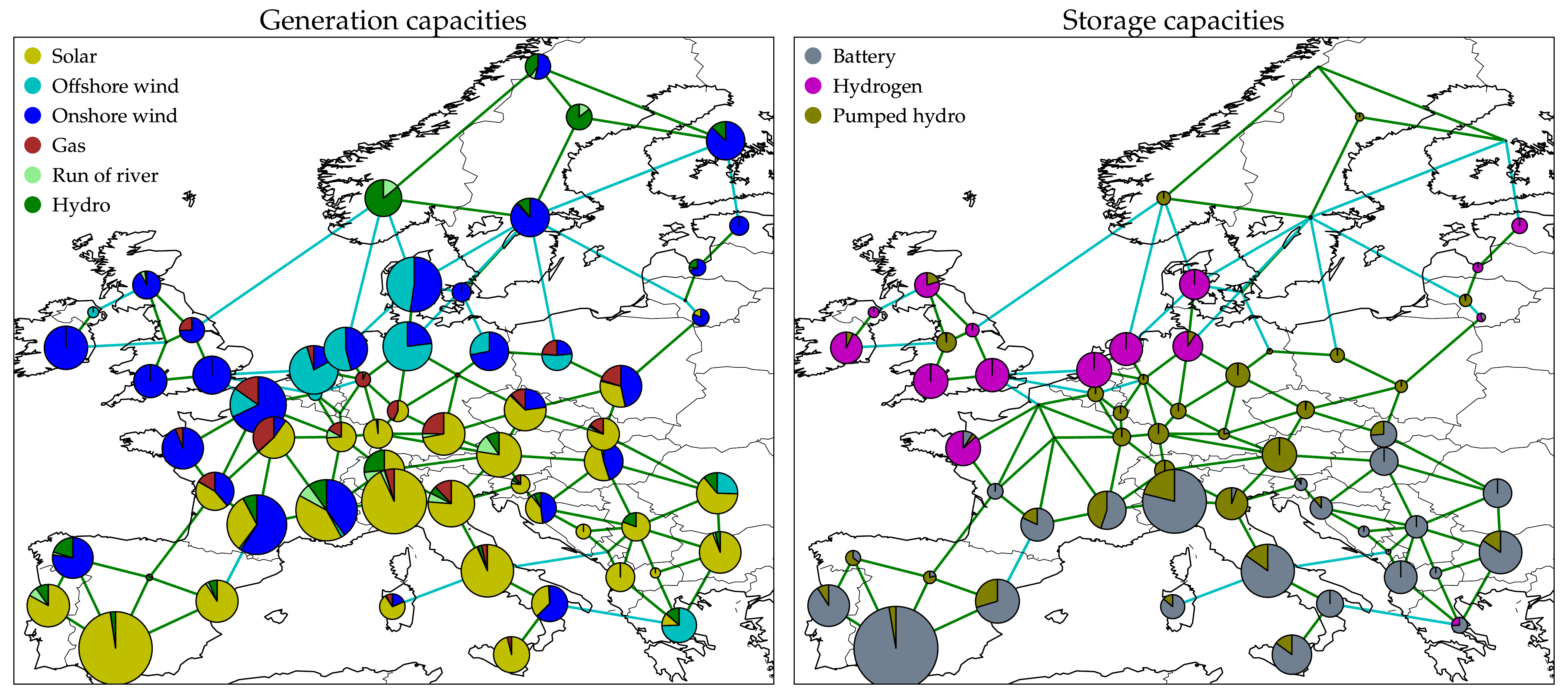}
\caption{Optimised generation layout (left) and storage layout (right) resulting from the optimisation \eqref{eq:objective}. AC lines in green and DC lines in blue.}
\label{fig:gen}
\end{figure*}

The input data for the system model is based on PyPSA-Eur, a dataset of the European electricity system containing spatially detailed information about the transmission network topology, conventional generators, hydro power, and time series for wind and solar power potential and demand, compiled from various sources~\cite{PyPSA, Hoersch2018b}. The 5612 transmission lines and 4653 substations within the dataset are merged using the k-means clustering algorithm to 64 nodes and 132 transmission lines covering 33 countries, see Figure~\ref{fig:gen}. The distribution of generation capacities as well as generation and load time series are also aggregated to yield corresponding nodal representations. We choose a spatial resolution of 64 nodes for the European system in order to work on a coarse-grained  level while still being able to capture patterns on regional scale within larger countries. See~\cite{Hoersch2017} for further details on the underlying data set, and in particular for a discussion of network aggregation methods and the role of spatial scale for electricity system modelling.

The model uses a techno-economic optimisation minimising total annual system costs:\begin{equation}\label{eq:objective}
\min_{\substack{G_{n}^{\alpha},F_l, \\ g_{n}^{\alpha}(t),f_{l}(t)}} \left[\sum_{n,\alpha}c_{n}^{\alpha}G_{n}^{\alpha} + \sum_{l}c_{l}F_{l} + w_{t}\sum_{n,\alpha,t}o_{n}^{\alpha}g_{n}^{\alpha}(t)\right]~.
\end{equation}
Here $G_{n}^{\alpha}$ are the capacities of generation and storage technologies $\alpha$ at node $n$ and their associated fixed costs $c_{n}^{\alpha}$, $g_{n}^{\alpha}(t)$ is the nodal dispatch during hour $t$ and the associated operating cost $o_{n}^{\alpha}$, and $F_l$ are the line capacities and their associated fixed costs $c_l$. The model is run using weather and demand data for a representative year chosen to be 2012. To keep computation time reasonable the model is run for every third hour of the representative year leading to the weighting $w_{t}=3$ in the objective function and following constraints. As fossil fuel generators we assume open cycle gas turbines, which are more flexible but less efficient than combined cycle gas turbines. Renewable generators include solar PV, onshore wind and offshore wind. Batteries and hydrogen storage are used as extendable storage options, whereas hydroelectricity capacities (run-of-river, reservoirs and pumped storage) are fixed to today's level. All cost assumptions are given in~\cite{Hoersch2017}.

For every (weighted) hour, the demand at each node $d_{n}(t)$ must be met by local generation and storage discharge or by imported power flows $f_{l}(t)$ on transmission line $l$,
\begin{equation}
\label{eq:nodal-balance}
\sum_{\alpha}g_{n}^{\alpha}(t)-d_{n}(t) = \sum_{l}K_{n,l}f_{l}(t)~,
\end{equation}
where $K_{n,l}$ is the incidence matrix representing Kirchhoff's Current Law. For the HVAC part of the network also Kirchhoff's Voltage Law is enforced by demanding that the voltage differences around any closed cycle must sum to zero~\cite{Hoersch2017,Hoersch2018c}.

The dispatch $g_{n}^{\alpha}(t)$ of conventional generators is constrained by their capacity, expressed by the condition $
0\leq g_{n}^{\alpha}(t)\leq G_{n}^{\alpha}
$. Similarly, the dispatch of renewable generators is constrained by their capacity, $
0 \leq g_{n}^{\alpha}(t)\leq \bar{g}_{n}^{\alpha}(t)G_{n}^{\alpha}
$, where $\bar{g}_{n}^{\alpha}(t)$ is the fraction of capacity available depending on the weather conditions obtained from historical reanalysis weather data. When the generation is less than the available energy the remainder is curtailed.

The state-of-charge of all storage facilities must be consistent across all hours:
\begin{align}
\label{eq:storage}
\begin{split}
\mathrm{soc}_{n}^{\alpha}(t) &= \mathrm{soc}_{n}^{\alpha}(t-1) + w_{t}g_{n,\mathrm{inflow}}^{\alpha}(t) - w_{t}g_{n,\mathrm{spillage}}^{\alpha}(t) \\
&\hspace{4mm} \pm \begin{cases}
w_t\eta_1g_{n}^{\alpha}(t), & \mathrm{charging}\\
w_t\eta_2^{-1}g_{n}^{\alpha}(t), & \mathrm{discharging}
\end{cases}~.
\end{split}
\end{align}
Here $\eta_1$ and $\eta_2$ are the charging and discharging efficiencies, respectively. These efficiencies ensure that the storage facilities are only charged when there is an oversupply of power, and discharged only when generators and imports are not able to fully serve the demand. The state of charge is limited by the storage energy capacity, $
0 \leq \mathrm{soc}_{n}^{\alpha}(t) \leq E_{n}^{\alpha}
$, which is defined by the nominal power $G_{n}^{\alpha}$ through
\begin{equation}\label{eq:storage-energy}
E_{n}^{\alpha} = h^{\alpha}_{\mathrm{max}}G_{n}^{\alpha}~.
\end{equation}
Here $h^{\alpha}_{\mathrm{max}}$ is the maximum number of hours that a storage facility can charge or discharge at the full nominal power. We set $h^{\alpha}_{\mathrm{max}} = 6\mathrm{h}$ for battery storage and pumped hydro, and $h^{\alpha}_{\mathrm{max}} = 168\mathrm{h}$ for hydrogen storage~\cite{Schlachtberger2017}. This constraint implies that there is no separate optimisation of storage power and energy capacity.

The power flow on the transmission lines is constrained by the transmission capacities, $
|f_{l}(t)| \leq F_l$. 
The sum of the product of transmission capacities and transmission line lengths are constrained by an overall maximum capacity
\begin{equation}
\label{eq:trans_constraint}
\sum_{l} l_lF_l \leq \mathrm{CAP}_\mathrm{trans},
\end{equation}
which is fixed to a moderate expansion corresponding to $150\%$ of the current transmission capacities
\begin{equation}
\mathrm{CAP}_\mathrm{trans} = 1.5 \cdot \sum_ll_lF_l^\mathrm{today}.
\end{equation}


CO$_2$ emissions are limited by a global constraint $\mathrm{CAP}_{\mathrm{CO}_2}$, defined by specific emissions $e_{\alpha}$ in CO$_2$-tonne-per-MWh of the fuel $\alpha$ and the efficiency $\eta_{\alpha}$ of the generator
\begin{equation}
\label{eq:co2_constraint}
\sum_{n,\alpha,t} \frac{1}{\eta_{\alpha}}w_{t}g_{n}^{\alpha}(t)e_{\alpha} \leq \mathrm{CAP}_{\mathrm{CO_2}}~.
\end{equation}
This constraint is set to a reduction of emissions of $95\%$ compared to 1990 levels. Since the only fossil fuel generators in the model are open cycle gas turbines, the constraint~\eqref{eq:co2_constraint} directly translates into the  amount of fuel burned by these generators, and thus into the amount of power generated from this source~\cite{Schlachtberger2017, Schlachtberger2018}.

\section{Flow tracing}\label{sec:ft}


For clarity we omit in this section the time index $t$ and decompose the generation term in the hourly nodal power balance~\eqref{eq:nodal-balance} into generators (conventional and renewable) $g_{n}^{\alpha}$, storage discharging $s_{n}^{\alpha,+}$, and storage charging $s_{n}^{\alpha,-}$. We define the net nodal inflow into the network and the net nodal outflow from the network as
\begin{align}
P_{n}^{\mathrm{in}} &= \max\left(
\sum_{\alpha}\left[g_{n}^{\alpha} + s_{n}^{\alpha,+} - s_{n}^{\alpha,-}\right] - d_n, 0
\right)~,\\
P_{n}^{\mathrm{out}} &= \max\left(
d_{n} - \sum_{\alpha}\left[g_{n}^{\alpha} + s_{n}^{\alpha,+} - s_{n}^{\alpha,-}\right] ,0
\right)~.
\end{align}
We rewrite~\eqref{eq:nodal-balance} as
\begin{equation}
P_{n}^{i\mathrm{n}} +  \sum_k f_{k\to n} = P_{n}^{\mathrm{out}} + \sum_{k}f_{n \to k}~,
\end{equation}
where the incidence matrix has been replaced with the sums of inflows and outflows. Assuming perfect mixing of the various flow components, the method of flow tracing follows the different nodal inflows $P_{m}^{\mathrm{in}}$ downstream through the network. The share $q_{n,m}$ of outflow from node $n$ (both through the network and into node $n$), which has been an inflow at node $m$ has to fulfil the following partial flow conservation:
\begin{equation}
\delta_{n,m}P_{n}^{\mathrm{in}} +  \sum_{k}q_{k,m} f_{k\to n} = q_{n,m}P_{n}^{\mathrm{out}} + \sum_{k}q_{n,m}f_{n \to k}~.
\end{equation}
Rearranging this equation yields the matrix equation formulation of flow tracing
\begin{align}
\label{eq:ft}
\delta_{n,m}P_{n}^{\mathrm{in}} = \sum_{k}\left[\delta_{n,k}\left(P_{n}^{\mathrm{out}} + \sum_{k'}f_{n\to k'}\right) - f_{k\to n}\right]q_{k,m}
\end{align}
which can be inverted to calculate the nodal mixes $q_{n,m}$~\cite{Hoersch2018}. We can now derive the fraction of load $d_{n}$ or storage charging $s_{n}^{\beta,-}$ associated with nodal inflow from generation technology $\alpha$ at node $m$. For this purpose we first define the internal mix for net exporters with $P_{n}^{\mathrm{in}} >0$:
\begin{equation}
r_{n,\alpha}^{\mathrm{in}} = \frac{
g_{n}^{\alpha} + s_{n}^{\alpha,+}
}{
\sum_{\alpha}\left(g_{n}^{\alpha}+s_{n}^{\alpha,+}\right)
}~.
\end{equation}
This internal mix is then attached to the nodal inflow from node $n$ and followed through the network, leading to the internal mix for net importers $n$ with $P_{n}^{\mathrm{\mathrm{out}}} >0$:
\begin{equation}
r_{n,(m,\alpha)}^{\mathrm{out}} = \frac{
\delta_{m,n}\left(g_{n}^{\alpha} + s_{n}^{\alpha,+} \right)+ q_{n,m}r^{\mathrm{in}}_{m,\alpha}P_{n}^{\mathrm{out}}
}{
\sum_{\alpha}\left(g_{n}^{\alpha}+s_{n}^{\alpha,+}\right) + P_{n}^{\mathrm{out}}
}~.
\end{equation}
The share of load and storage charging at node $n$ associated with generation or storage discharging from technology type $\alpha$ at node $m$ is then
\begin{equation}
\label{eq:shares1}
d_{n}(m,\alpha) = \delta_{n,m}r_{n,\alpha}^{\mathrm{in}}d_{n} \: , \ \
s_{n}^{\beta,-}(m,\alpha) = \delta_{n,m}r_{n,\alpha}^{\mathrm{in}}s_{n}^{\beta,-}
\end{equation}
for net exporters, and
\begin{equation}
\label{eq:shares2}
d_{n}(m,\alpha) = r_{n,(m,\alpha)}^{\mathrm{out}}d_{n} \: , \ \
s_{n}^{\beta,-}(m,\alpha) = r_{n,(m,\alpha)}^{\mathrm{out}}s_{n}^{\beta,-}
\end{equation}
for net importers. The scheme applied here assumes that first all inflow and outflow inside a node is aggregated, and then this aggregated flow is coupled to the network. This description is suitable for a coarse-grained system representation as used for this contribution. For a spatially more detailed representation, alternatively all inflows and outflows could be directly coupled to the network. The influence of choosing either approach on the flow tracing results will be studied in a forthcoming publication.


\section{Results}\label{sec:res}
The distribution of generation capacities in the scenario resulting from the system optimisation~\eqref{eq:objective} is shown in the left panel of Figure~\ref{fig:gen}. This layout is sensitive to the input parameters and optimisation constraints, in particular to the cap on the total transmission capacities and CO$_2$ emissions in~\eqref{eq:trans_constraint} and~\eqref{eq:co2_constraint}~\cite{Schlachtberger2018}. 
The figure shows that solar generation capacities are predominantly located in the southern half of the system, in line with the favourable solar radiation conditions in the southern countries. Offshore wind is located mainly at the North Sea and Baltic Sea along with minor capacities in the Black Sea and the Mediterranean Sea. Onshore wind is spread more evenly throughout the northern and western countries.

The right panel of Figure~\ref{fig:gen} displays the distribution of the optimised nominal power for storage technologies $G_{n}^{\alpha}$ in~\eqref{eq:storage-energy}. This corresponds to the ability of the storage facilities to balance hourly fluctuations in electricity production and demand. The total energy capacity can be calculated from~\eqref{eq:storage-energy}. Scaling the nominal power with $h^{\alpha}_{\mathrm{max}}$ shows that the energy capacity is largest for hydrogen storage followed by battery storage and last pumped hydro storage.

The spatial distribution of the generation and storage capacities proposes that the short term battery storage is paired with solar generation capacity to balance the strong diurnal pattern of solar power generation, whereas the long term hydrogen storage is associated with wind power generation capacity to balance weekly and seasonal weather patterns.

\begin{figure}[t]
\centering
\includegraphics[width=\linewidth]{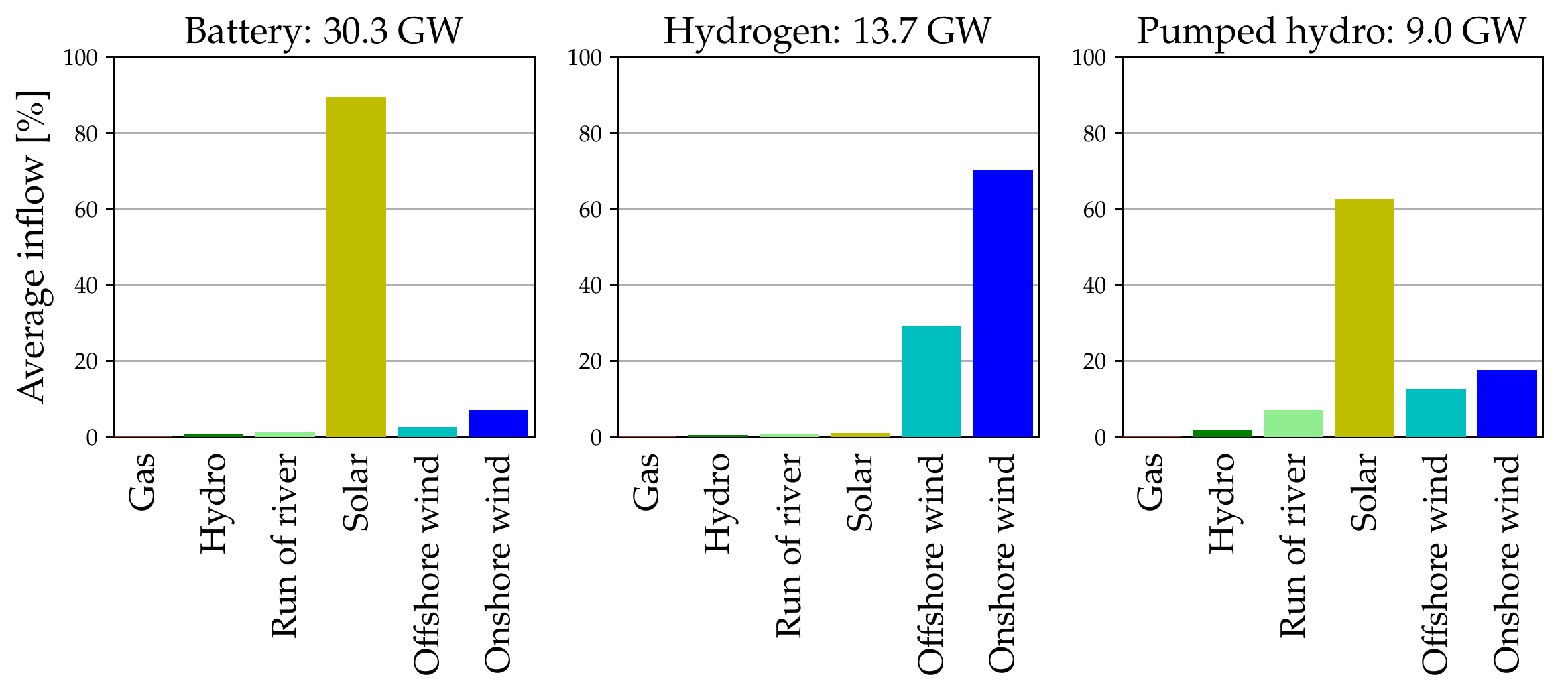}
\caption{Average hourly inflow per storage technology decomposed into the six generation technologies. The total average inflow is denoted in the title.}
\label{fig:stor}
\end{figure}

This intuition is confirmed by tracing the composition of power inflow for the charging of the different storage technologies. Figure~\ref{fig:stor} shows aggregated average hourly inflows $\langle \sum_{n,m}s_{n}^{\beta,-}(m,\alpha)\rangle$ for the three storage technologies for each of the six generation technologies. Corresponding with the spatial distribution shown in Figure~\ref{fig:gen}, hydrogen storage is mainly utilised for wind power generation, whereas battery storage mostly receives inflow from solar power generation. The pumped hydro storage capacities are fixed to today's layout, leading to a mixed utilisation of different generation technologies, but dominated by solar.

\begin{figure}[t]
\centering
\includegraphics[width=\linewidth]{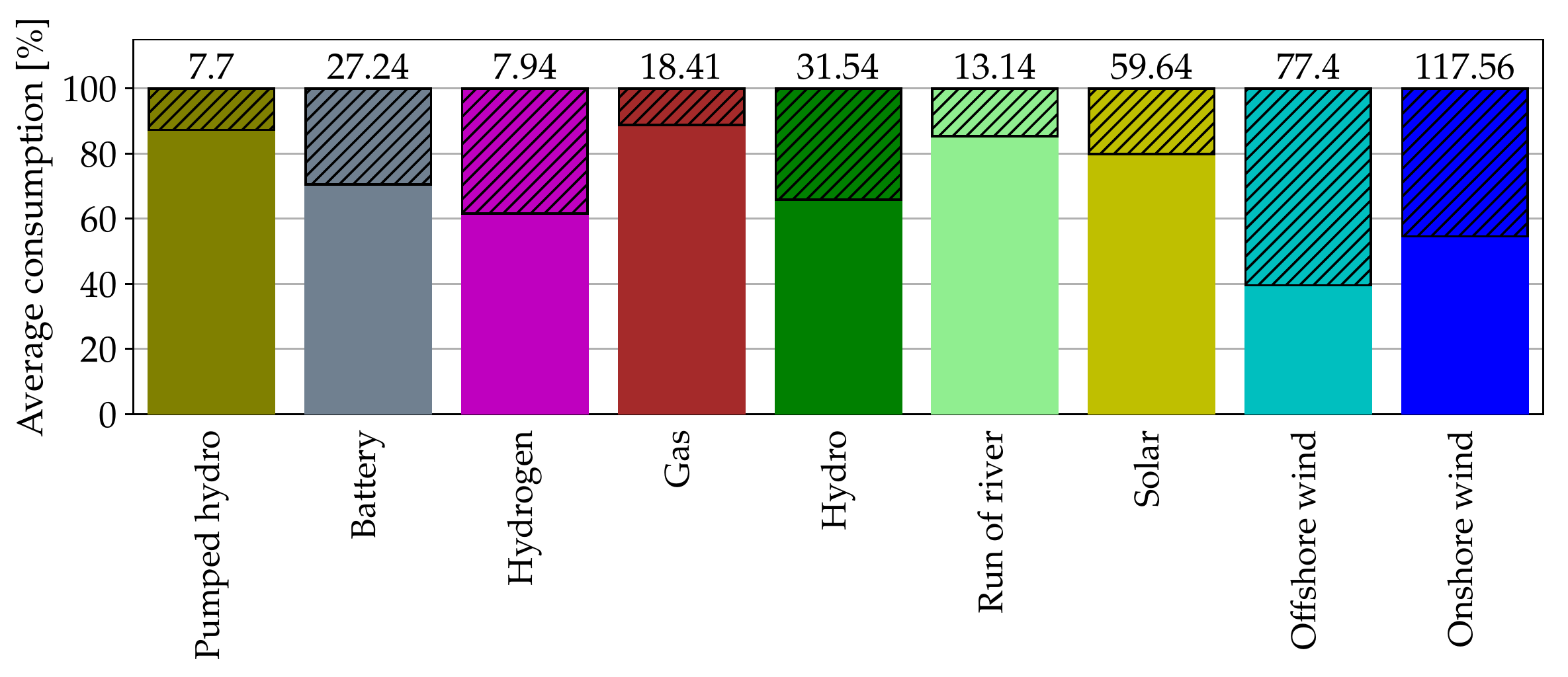}
\caption{Average local vs. non-local consumption per technology split between the source node and externally (shaded area). The absolute average consumption for each technology is written in GW above each bar.}
\label{fig:cons}
\end{figure}

Tracing the power flow originating from generation and storage technologies allows to assess how locally this power is consumed. Figure~\ref{fig:cons} shows how much of the average generation or storage discharging is consumed inside the same node, or alternatively is transmitted as a power flow over the network for consumption in another node (shaded area). Using the expressions in~\eqref{eq:shares1} and~\eqref{eq:shares2} this can be expressed as $\langle \delta_{n,m}d_{n}(m,\alpha)\rangle$ and $\langle \sum_{m\neq n}d_{n}(m,\alpha)\rangle$, respectively. The absolute average consumption associated with each generation or storage discharge is noted in GW above each bar. Note that due to the efficiencies in~\eqref{eq:storage} the total storage discharging is lower than the total storage charging as given in Figure~\ref{fig:stor}.

We observe that pumped hydro discharging, gas and run of river power generation is predominantly consumed locally (87$\%$, 89$\%$ and 85$\%$, respectively). While the placement of pumped hydro and run-of-river capacities is not optimised in the system, the cost structure of open cycle gas turbines (low capital costs, high marginal costs) proposes that this technology is locally deployed for peak demand covering when other flexibility options are not cost optimal. Solar and hydro power generation are also mostly consumed locally ($80\%$ and $66\%$, respectively), since they represent the predominant local generation capacities when they are installed in a node. Nevertheless, for battery and hydrogen storage the $30\%$ to $38\%$ of non-local usage show that the system uses these storage technologies as a system-wide backup. Wind power generation is consumed locally up to $55\%$ (onshore) and $40\%$ (offshore), but due to its massive deployment in the system it is also to a comparatively higher share transmitted over the network for consumption at other nodes. These results also reflect the spatial distribution of renewable generation resources for wind and solar power PV: whereas favourable wind power conditions occur in general only distant from load centres, solar PV is less locationally sensitive within each country and can be built close to the loads. Note that for simplicity we do not use the ability to inter-temporarily trace power flow through the storage and discard the information of the original source of the storage outflow.

\begin{figure}[t]
\centering
\includegraphics[width=\linewidth]{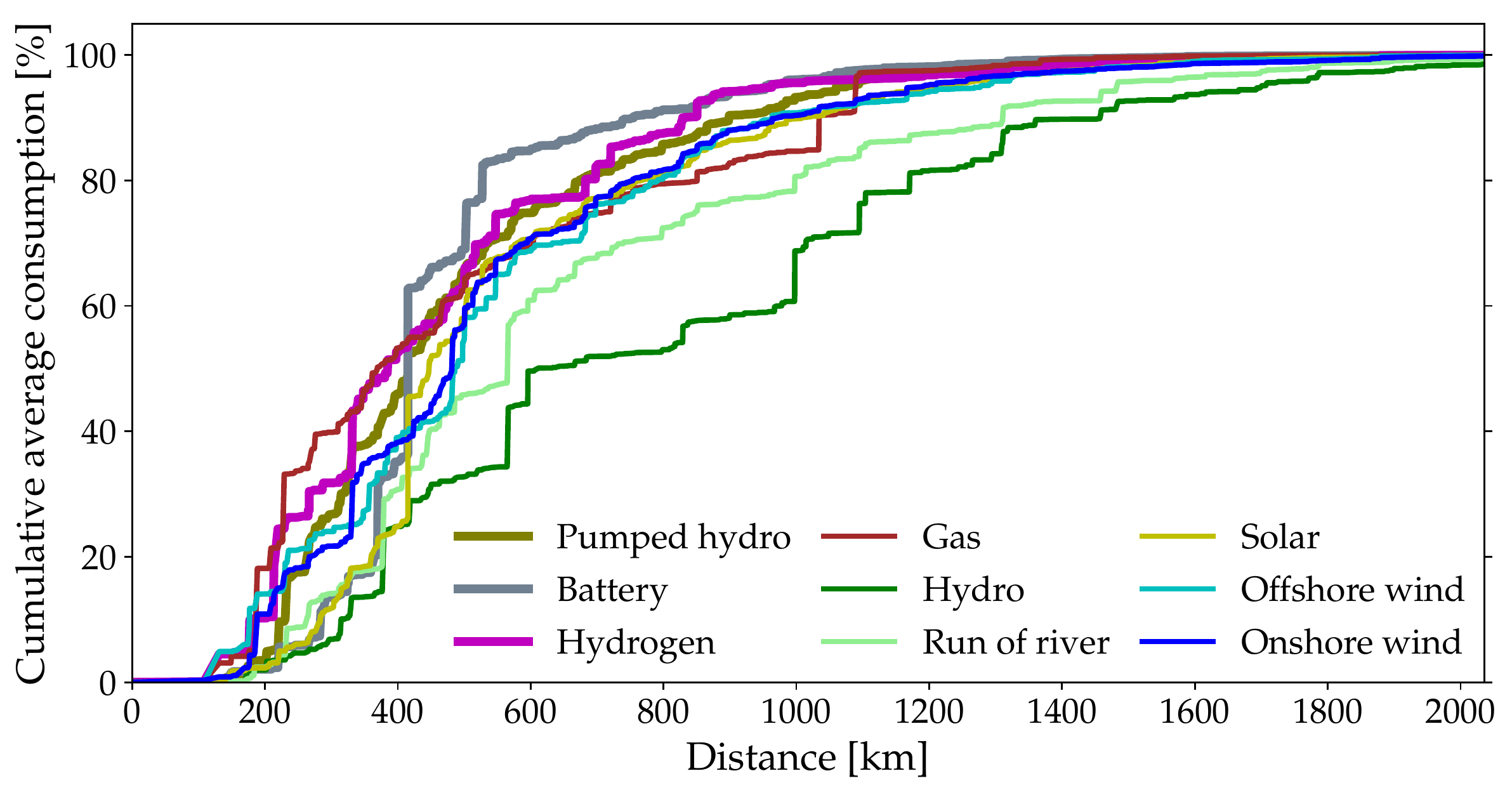}
\caption{Cumulative average consumption as a function of distance. Only exported power is included hence the starting point at zero. The lines in this figure correspond to the shaded areas in Figure~\ref{fig:cons}.}
\label{fig:distance}
\end{figure}

Figure~\ref{fig:distance} shows the cumulative average consumption per technology as a function of the spatial distance between the exporting and consuming node. The figure is cut off at 2036km at which point all technologies have reached 99\%. For reference, the largest distance between two nodes in the network is 3455km. The lines in this figure correspond to the shaded areas in Figure~\ref{fig:cons}. The three storage technologies (highlighted with thicker lines) have a tendency to be consumed more locally than the generation technologies. Most of the generation and storage technologies follow a similar pattern, except reservoir hydro, and, to a lesser extent, run of river, which are both being exported over large distances in the network. The results in this figure emphasize the importance of transmission capacity for the system. Although the power from most generation and storage technologies on average is consumed predominantly locally, in case of exports it is often distributed over wide parts of the system.

\section{Discussion \& Conclusions}\label{sec:con}
In this article we have extended the application of flow tracing to power flows associated with charging and discharging of storage capacities in a low-carbon scenario of a future European electricity system first presented in~\cite{Hoersch2017}. Using this tracing approach we are able to determine the composition of storage inflow with respect to the different generation technologies present in the system. We observe that short-term battery storage is predominantly used by solar power generation, whereas longer-term hydrogen storage is almost completely charged with power from onshore and offshore wind power generation. This flow-based result quantitatively confirms the intuition gained from the spatial distribution of generation and storage capacities. Bearing in mind the limits of the spatial resolution of 64 nodes for the European electricity system we furthermore determine how much of power generation or storage outflow is consumed in the same node or alternatively distributed over the transmission network to loads at other nodes in the system. It is shown that storage outflow is mostly consumed locally inside the same node. This is similar for the usage of hydro or solar PV power generation, with the local usage even more pronounced for power from open cycle gas turbines. In contrast, power generation from onshore wind is less locally consumed, and offshore wind power is predominantly exported to other nodes in the network. Whereas these findings propose an interpretation of a generally more local usage of storage technologies, the analysis of the exported power flows show that these often stretch across large parts of the system. For all generation and storage technologies around $20\%$ of the average exported power flow is consumed at nodes which are more than 1000 km away from the location of network inflow. Our flow-based analysis thus suggests a local-but-global usage of storage capacities -- whereas on average these capacities are deployed locally, if needed their flexibility is used also by distant nodes connected through sufficient transmission capacities of the power grid.

The study presented in this contribution calls for an extension in several directions. Increasing the spatial resolution of the system representation and considering different levels of transmission expansion allows a more detailed investigation of the local-but-global usage of storage capacities in a low-carbon European electricity system. Analysing the time-series of corresponding flow patterns will further shed light on the system conditions which correspond to either a local or global impact of different generation and storage technologies. Combining this information with a flow-based nodal cost allocation mechanism could inspire new economic contract concepts for future electricity markets with a high share of renewable generation. Equally important, by revealing the details of the system benefit of power transmission, the flow-based system analysis as advocated in the presented analysis is a valuable contribution in the context of public discussions on transmission expansion.

\section*{Acknowledgements}
M. S. is funded by The Carlsberg Foundation Distinguished Postdoctoral Fellowship. 
M. G. and T. B. are partially funded by the \href{https://reinvestproject.eu/}{RE-INVEST project}
, which is supported by Innovation Fund Denmark (6154-00022B). T. B. acknowledges funding from the Helmholtz Association under grant no. VH-NG-1352. The responsibility for the contents lies solely with the authors.



%

%
%

\bibliographystyle{IEEEtran}
\bibliography{references}

\end{document}